\documentclass[prl,aps,twocolumn,showpacs]{revtex4}
\usepackage{graphicx,psfrag}

\newcommand{\A}{{\rm A}}
\newcommand{\B}{{\rm B}}
\newcommand{\C}{{\rm C}}

\begin{document}
\title{Artificial 
electromagnetism for neutral atoms: Escher staircase and Laughlin liquids}
\author{Erich J. Mueller}
\affiliation{Laboratory of Atomic and Solid State Physics, Cornell University, Ithaca, New York 14853
}
\date{\today}

\begin{abstract}
We show how lasers may create fields
which couple to neutral atoms in the same way that the electromagnetic fields couple to charged particles.  These fields are needed for using neutral atoms as an {\em analog quantum computer} for simulating the properties of many-body systems of charged particles.  They allow for seemingly paradoxical geometries, such as a ring where atoms continuously reduce their potential energy while moving in a closed path.   We propose neutral atom experiments which probe quantum Hall effects and the interplay between magnetic fields and periodic potentials.
\end{abstract}
\pacs{03.75.-b, 
03.75.Lm, 
03.67.Lx, 
32.80.Lg, 
73.43.-f, 
}
\maketitle

Recently, many researchers have expressed interest in using ultracold alkali atoms as 
{\em analog quantum computers} to simulate properties of solid state systems \cite{simulation}.  For example, the leading model of high temperature superconductivity, the Hubbard model, can be studied by placing alkali atoms in an {\em optical lattice} -- a periodic potential formed by interfering several laser beams.
Experimental realizations of the Hubbard model could show whether it captures the phenomena of high temperature superconductivity.  
Similarly,
cold gases provide an ideal setting for studying models of  quantum Hall effects \cite{bosonqh} and exotic phase transitions \cite{exotic}.

%
%


A major impediment to studying some of these models, such as those describing quantum Hall effects, 
%
is the lack of fields which couple to the neutral atoms in the same way that the electric and magnetic fields couple to charged particles.  Here, we show how to create these {\em artificial} electromagnetic fields.  Since these fields are only analogies of the real electric and magnetic field
they do not obey Maxwell's equations.  One can therefore create {\em unphysical} and counterintuitive field configurations which lead to a set of as-yet unstudied behavior.  Among our examples of these seemingly {\em impossible} field configurations, we describe an `Escher staircase' setup where atoms can move around a closed path, continually reducing their potential energy.

The literature already contains several, somewhat limited, implementation of electrical and magnetic fields for neutral atoms.  Experimentalists routinely use the Earth's gravitational field as an analog of a uniform electric field \cite{gravelectric}.  They also study  systems in non-inertial frames:  uniform acceleration is equivalent to a constant electric field \cite{acceleratinglattices}, while circular motion corresponds to a uniform magnetic field \cite{rotating}.  Recently, Jaksch and Zoller \cite{zoller} described a method where an effective magnetic field can be applied to two-state atoms in an appropriately designed optical lattice in the presence of an external `electric field'.  Our approach is an elaboration of Jaksch's, where the two-state atoms are replaced by three-state atoms.  This allows us to overcome the major deficiency of Jaksch and Zoller's scheme:   we do not need an external electric field to generate the magnetic field.
This improvement comes at the cost of more complicated laser configurations.

As in these prior analogs of electromagnetism, our  artificial fields contain no dynamical degrees of freedom.  Therefore they neither give rise to analogs of Coulomb interactions between the neutral atoms, nor do they support analogs of light.

Subsequent to the preparation of this manuscript,  another scheme for generating analogs of electromagnetic fields was suggested by Sorensen, Demler, and Lukin \cite{sorensen}.  That work uses time dependent hopping matrix elements along with a large oscillating quadrupolar potential.  Compared to our approach, Sorensen et al.  use a much simpler laser configuration, however there are nontrivial technical issues involved with the stability of the oscillating potential.

{\bf Basic Setup:}
Our approach relies upon creating an optical lattice with three distinct sets of minima.  Each of these minima trap a different internal state of the neutral atoms.  The internal states will be labeled '$\A$', '$\B$', and '$\C$', and the minima will be labeled by their location and by the state that is trapped at that location.  For example, figure~1(a) shows a one-dimensional array labeled as $\cdots$-$\A_1$-$\B_2$-$\C_3$-$\A_4$-$\B_5$-$\C_6$-$\A_7$-$\B_8$-$\cdots$.  Importantly, this setting breaks parity symmetry.

\begin{figure}
\includegraphics[width=\columnwidth]{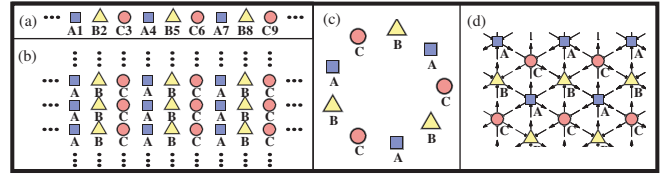}
\vspace{-0.3in}
\caption{(Color Online)  Lattices of three types of sites: (a) 1-D chain, (b) square, (c) ring, and (d) triangular.}
\end{figure}

Looking at this one dimensional chain,
an atom in state $\A$, sitting in site $\A_4$, is immobile.  The atom cannot hop to site $\B_5$ or $\C_3$, because it would need some mechanism for changing its internal state.  The probability of tunneling by three sites to $\A_1$ or $\A_7$ is astronomically small.

We turn on hopping between site $\A_4$ and $\B_5$ by introducing a laser  with the following properties:  (i) the laser frequency $\omega_{AB}$ is close to the energy differenced between the internal states $\A$ and $\B$ (ie. $\omega_{AB}\sim E_\A-E_\B$); (ii) the laser polarization is chosen so that the transition from internal state $\A$ to $\B$ is allowed; (iii)  the laser cannot induce transitions from states $\A$ to $\C$ or from $\B$ to $\C$, either because the transition is forbidden, or because the detuning is too great.  One does not have to use a single laser to drive this transition but can instead use a Raman transition, which involves multiple lasers and the virtual occupation of one or more intermediate state.

In the presence of this laser field, the atom can explore a two state Hilbert space.  In the rotating wave approximation, the time dependent Schroedinger equation is
\begin{equation}\label{matham}
\textstyle
\begin{array}{c}
\textstyle i\partial_t \left(
\begin{array}{c} 
\psi_{\A4}\\
\psi_{\B5}
\end{array}
\right)
=
H(t) \left(
\begin{array}{c} 
\psi_{\A4}\\
\psi_{\B5}
\end{array}
\right)\\[4mm]
H(t)=\left(\begin{array}{cc}
E_\A&-\Omega_{\A\B}e^{-i(\omega_{\A\B}t+\phi)}\\
-\Omega_{\A\B}e^{i(\omega_{\A\B}t+\phi)}&E_\B
\end{array}\right).
\end{array}
\end{equation}
The quantum mechanical amplitude for the particle being in state $\A$ ($\B$) on site $\A_4$ ($\B_5$) is $\psi_{A4}$ ($\psi_{\B5}$).
The energy of the internal states $\A/\B$ are $E_{\A/\B}$.  The Rabi frequency $\Omega_{\A\B}$ is proportional to the product of the laser amplitude and the overlap between the states trapped in $\A_4$ and $\B_5$.    We take $\Omega_{\A\B}$ to be real, and introduce a phase $\phi$, which is related to the phase of the coupling laser.   
In particular, if we translated the entire lattice by some distance $\bf r$, the phase $\phi$ would change by $\phi\to\phi+{\bf q\cdot r}$, where $\bf q$ is the wave-vector of the coupling laser \cite{raman}. 

This, and future Hamiltonians are more compactly written in a second quantized notation, 
\begin{eqnarray}
H&=&E_\A \hat\psi_{\A4}^\dagger \hat\psi_{\A4}
+E_\B \hat\psi_{\B5}^\dagger \hat\psi_{\B5}\\\nonumber&&
-\Omega_{\A\B}\left(e^{-i(\omega_{\A\B}t+\phi)}
\hat\psi_{\A4}^\dagger \hat\psi_{\B5}
+e^{i(\omega_{\A\B}t+\phi_{\A})}
\hat\psi_{\B5}^\dagger \hat\psi_{\A4}\right)
\end{eqnarray}
where, for example, creation and annihilation operators $\hat\psi_{\A4}^\dagger$ and $\hat\psi_{\A4}$ add and remove a particle from site $\A_4$ in internal state $\A$.  In the non-interacting system, the operators $\hat \psi$ obey the same equations of motion as the wave-function $\psi$ in (\ref{matham}).  At the single-particle level it does not matter whether we use bosonic or fermionic commutation relations. Where no confusion will result, we may neglect the letter $\A$ which denotes the internal state.

We apply time-dependent canonical transformations of the form $\hat\psi_j\to e^{i f(t)}\hat\psi_j$, $\hat\psi_j^\dagger\to e^{-i f(t)}\hat\psi_j^\dagger$.  As is readily verified from the equations of motion (\ref{matham}), under this transformation the Hamiltonian becomes
$H\to H-f^\prime(t) \hat\psi_j^\dagger \hat\psi_j$.  In particular we can construct a time independent Hamiltonian by transforming into the `rotating frame,'
\begin{eqnarray}
\hat\psi_{\A4}&\to& e^{-i(E_\A t-\phi)}\hat\psi_{\A4}\\
\hat\psi_{\B5}&\to& e^{-i(E_\B t+\Delta_{\A\B})}\hat\psi_{\B5}\\
H&=&-\tau (\hat\psi_{4}^\dagger \psi_{5}+ \hat\psi_{5}^\dagger \psi_{4})
+\Delta \psi_5^\dagger\psi_5,
\end{eqnarray}
where $\tau=\Omega_{\A\B}$ and  $\Delta=\omega_{\A\B}-(E_\A-E_\B)$.

Introducing two more lasers, coupling states $\B$-$\C$, and $\C$-$\A$ with appropriately chosen intensities and detunings, this same procedure yields the Hamiltonian
\begin{equation}
H=\textstyle\sum_j\left( j \Delta (\hat\psi_j^\dagger\hat\psi_j)-\tau
(\hat\psi_j^\dagger\hat\psi_{j+1}+\hat\psi_{j+1}^\dagger\hat\psi_j)
\right),
\end{equation}
corresponding to a one-dimensional chain of sites in a uniform electric field.  As is shown below, this same approach can produce electric fields in higher dimensions.  In this case, momentum transfer from the lasers will generate an effective magnetic field.

{\bf Higher Dimensions:}
In more complicated geometries there may not be a Canonical transformation which leads to a time independent Hamiltonian.  However, the time dependence takes a simple form if one transforms
\begin{equation}
\hat\psi_{\mu_j j} \to e^{-iE_{\mu_j }t} \hat\psi_{\mu_j j},
\end{equation}
where $j$ labels the site located at $\bf r_j$, and $\mu_j=\A,\B,\C$ gives the internal state which is trapped at that site.  The Hamiltonian then becomes
\begin{equation}\label{hop}
H=-\sum_{\langle i j\rangle} \tau_{\mu_i\mu_j}\left(e^{i {\bf q}_{\mu_i\mu_j}\cdot {\bf R_{ij}} }e^{-i\Delta_{\mu_i\mu_j}t}
\psi_{\mu_ii}^\dagger \psi_{\mu_jj}+{\rm H.C.}\right).
\end{equation}
The sum includes all nearest neighbor sites $\langle ij\rangle$.  The internal state trapped at site $i$ is $\mu_i$.  The bond position is  ${\bf R_{ij}}=({\bf r_i+r_j})/2$.  The hopping is $\tau_{\mu\nu}=\Omega_{\mu\nu}$ for $\mu\neq\nu$, and
$\tau_{\mu\mu}=\tau_0$.  The parameter $\tau_0$ is given by the overlap of the wavefunctions  on neighboring sites.  The wave-vector of the laser coupling state $\mu$ to $\nu$ is ${\bf q_{\mu\nu}}$ (so $\bf q_{\mu\mu}=0$).  The detuning is $\Delta_{\mu\nu}=\omega_{\mu\nu}-(E_\mu-E_\nu)$ when $\mu\neq\nu$, and $\Delta_{\mu\mu}=0$.  The letters ${\rm H.C.}$ denote the Hermitian conjugate of the previous term.

If all of the laser intensities are adjusted so that $\tau_{\mu\nu}=\tau_0$ for all $\mu,\nu$, then equation (\ref{hop}) is formally the equation of motion of a particle with charge $e$ in a vector potential defined on the bonds by
\begin{equation}\label{vecpot}
\frac{e}{c}{\bf A(R_{ij})\cdot r_{ij}} ={\bf q}_{\mu_i\mu_j}\cdot {\bf R_{ij}}-\Delta_{\mu_i\mu_j} t,
\end{equation}
where ${\bf r_{ij}=r_i-r_j}$.  

Using this mapping to a vector potential, we can construct many interesting field configurations.  For example, consider a lattice with 
 the striped geometry shown in figure~1(b), where as one moves in the $\bf\hat x$ direction, one encounters alternating rows of sites $\A$, $\B$, and $\C$.    With this geometry, only the x-component of the vector potential, $A_x$, will be non-zero.  In the simplest case, where each of the three coupling lasers have the same wave-vector $\bf q$ and detuning $\Delta$, the vector potential is ${\bf A(r)}={\bf \hat x}(c/ed)({\bf q\cdot r}-\Delta t)$, where $d$ is the lattice spacing.  This corresponds to a uniform electric field ${\bf E} = -{\bf \hat x} \Delta c/ed$ and a uniform magnetic field ${\bf B}= {\bf \hat x\times q} (c/ed)$.  By changing the relative angle between $\bf q$ and the $\bf\hat x$ axis, one can control the strength of the magnetic field.  Since the recoil momentum $q$ can be made comparable to the inverse lattice spacing, one should be able to construct extremely large fields where flux through a unit cell of the lattice exceeds the magnetic flux quantum $\Phi_0=2\pi c/e$.

If $q$ is aligned with the hopping direction, then the effective magnetic field vanishes, resulting in an electric field without a magnetic field.

{\bf Applications:}
Earlier we introduced some interesting problems which could be addressed by applying effective electric and magnetic fields to a system of particles on a lattice.  Here we discuss a further
possibilities.

At moderate values of the ``magnetic field" experiments could explore how the periodic potential affects vortex structures in a Bose condensate \cite{duine}.  
One could also study vortex physics near ``pairing transitions" where the structure of vortices change \cite{pairing}

At much larger fields ($\Phi\sim\Phi_0$) Jaksch and Zoller \cite{zoller} recently discussed the exciting idea of using neutral atoms to study the fractal energy spectrum that Hofstadter \cite{hofstadter} predicted  for noninteracting charged particles on a lattice in a magnetic field.  
The spectral gaps would be observable as plateaus in the density of noninteracting harmonically trapped fermions.
It would be even more exciting to explore an interacting system in this same regime, and study fractional quantum Hall physics, and the interplay between quantum Hall effects, Mott insulating physics, and this fractal single-particle spectrum \cite{qhlat}.  The simplest such experiment would use the geometry in figure~1(b), and allow the system to equilibrate with $\Delta=0$.  All single particle observables are measurable through imaging, while photoassociation provides access to the short range pair correlation function\cite{bosonqh}.  Some transport measurements are possible by detuning the lasers so that $\Delta\neq0$.

Several authors have shown that
for filling fractions $1/2<\nu<6$, bosons with short range interactions in a strong magnetic field will form non-trivial many-body states \cite{bosonqh}.  Fermions are more tricky, as s-wave interactions (which dominate at low temperatures) cannot lead to fractional quantum Hall effects in fermions.  However, resonantly enhanced p-wave interactions can lead to such correlated states \cite{regnault}.

Previous proposals for creating analogs of quantum Hall states in cold atoms relied upon rotation to provide the effective vector potential.  Such schemes are made difficult by the need to carefully balance the centripetal force which maintains rotation and the harmonic trapping potential.  The window of rotation speeds for finding strongly-correlated physics falls off with the inverse of the number of particles.  The present approach does not require this delicate balancing of forces, and therefore allows one to study these effects in a macroscopic system.

Not only are magnetic fields of interest, but so are large electric fields.  For example Sachdev et al. \cite{sachdev} have discussed the intricate Mott-Insulator states which are found when the `voltage difference' between neighboring wells is comparable to the on-site repulsion.  The method presented here is a powerful tool for studying such states.


{\bf Unphysical Fields:}
We once again emphasize that although $\bf A$ couples to the neutral atoms as if it were a vector potential, it does not obey Maxwell's equations.  Consequently, one can engineer seemingly paradoxical geometries.  Consider, for instance, the ring of sites illustrated in figure~1c, with all detunings set equal.    According to equation~(\ref{vecpot}), there is a uniform `electric' field pointing along the chain.  Thus a particle can move around the ring, continuously moving to a lower potential energy, returning to the starting point, but (by conservation of energy) having gained a great deal of kinetic energy.  One can repeat the process {\em ad infinitum}; the maximum velocity is limited only by Umklapp processes.
That is, when the particles deBroglie wavevector  coincides with the intersite distance, the matter-wave is Bragg reflected off of the lattice, and reverses direction.  If the chain was not bent in a circle, this reflection would lead to the familiar Bloch oscillations.  No conservation laws are violated by this continuous acceleration, as the lasers provide a source of energy and momentum.

This bizarre situation where a particle can reduce its potential energy by moving in a closed path is reminiscent of the optical illusion in MC Escher's print ``Ascending and Descending,"   where a staircase forms a continuously descending closed loop.  The quantum mechanical properties of a particle in such a chain of $N$ sites are ascertained by noting that the Hamiltonian, $H=-\tau \sum_{j=1}^N( e^{i\delta t } \psi_j^\dagger \psi_{j+1}+e^{-i\delta t} \psi_{j+1}^\dagger \psi_{j}),$ with $\psi_{N+1}\equiv \psi_1$, is translationally invariant, and therefore extraordinarily simple in momentum space.  In terms of operators $a_k=\sum_j e^{-2\pi i j k/N}\psi_j/\sqrt{N}$, the Hamiltonian is diagonal, $H=\sum_k E_k(t) a_k^\dagger a_k.$  The eigenvalues $E_k(t)=-2 \tau \cos(2\pi k/N+\delta t)$ are time dependent, reflecting the non-equilibrium nature of the system.  The motion of a wave-packet is determined by the instantaneous phase velocity 
\begin{eqnarray}
v &=& \frac{d N}{2\pi} \frac{\partial E_k}{\partial k} = 2 \tau d \sin(2\pi k/N + \delta t),
\end{eqnarray}
which oscillates as a function of time.  The factor of $dN/2\pi$, where $d$ is the intersite spacing, converts the velocity into physical units.  This oscillation is exactly the Bragg diffraction previously mentioned.  During one period of oscillation, the particle moves around the ring approximately $2 \tau/(N\delta)$ times. 

A more complicated geometry with similar paradoxical properties is illustrated in figure~1d.  In this structure, a triangular lattice is formed from three interpenetrating sublattices with wells of type $A$, $B$, and $C$.  Here, a constant detuning yields a very intricate `unphysical' electric field configuration: arrows depict 
directions in which hopping reduces the potential energy.
 Upon traversing alternate plaquettes, a particle can continuously increase, or decrease its potential energy.  To understand the behavior of a particle in this lattice, one once again relies upon translational invariance, and introduces operators $a_k=\sum_{\bf r} \psi_{\bf r} e^{-ik\cdot r}$, where $k$ lies in the first Brillouin zone (BZ) of the triangular lattice, and the sum is over all lattice sites.  The Hamiltonian is then
\begin{eqnarray}
H &=& -\tau \sum_r \left[e^{i\delta t} \left({\textstyle\sum_{j=1}^3 \psi_r^\dagger \psi_{r+r_j}}
\right)+{\rm H.C.}\right]\\
&=&-2\tau \int_{BZ}\frac{d^2k}{\Omega} a_k^\dagger a_k \textstyle \sum_{j=1}^3 \cos({\bf k\cdot r_j}+\delta t) .
\end{eqnarray}
The lattice generators $\{{\bf r_1,r_2,r_3}\}$ connect nearest neighbor sites, and are illustrated by arrows in figure~1d.  Only two of these generators are linearly independent ($\bf r_1+r_2+r_3=0$).  The area of the first Brillouin zone is $\Omega=8\pi^2/\sqrt{3} d^2$, where $d$ is the lattice spacing.  Again, the group velocity of a wave packet is simply the gradient of the energy $E_k=-2\tau \sum_j \cos({\bf k\cdot r_j}+\delta t)$.  Of particular note is the fact that at the zone center ($k=0$) the group velocity is alway zero.  Thus a stationary packet remains stationary.  This result is not surprising, since there is nothing in the geometry which picks out a direction in which the packet could start to move.

More surprising is the fact that the effective mass, related to the curvature of $E_k$ is oscillatory at $k=0$, spending equal amounts of time positive and negative.  When the effective mass is negative, quantum diffusion acts opposite to its normal behavior, and wave packets become sharper.  Thus localization occurs: the wave packet's size oscillates periodically, rather than continually growing.  Similarly, if the packet has a small momentum with $|k|\ll 2\pi/d$, then the particle does not simply propagate ballistically, but its velocity oscillates sinusoidally about $v=0$, and the particle is trapped near its initial location.

{\bf Physical Realization:}
There are many ways to engineer the three-state lattices described above.  The difficult task is to produce the confinement and Raman couplings with a small number of lasers in a geometry which can be easily implemented.
A detailed analysis of the various configurations goes beyond the scope of this paper, and a more comprehensive article is in preparation.

A key idea is that if the internal states are related by symmetries (ex. a spin-1 multiplet), then the various traps can be created by the same lasers, and the ($A$-$B$) and ($B$-$C$)  Raman transitions can use the same drive.  Driving transitions with microwave or RF fields, rather than lasers, will reduce the need for optical access \cite{rf}.  

An alternative approach is to note that one can create analogs of electromagnetic fields even if the sites $A$, $B$, and $C$, trap atoms in the same state.  One can instead rely on a superlattice structure, where the energies of the three sites differ by large amounts \cite{superlattice}.
Hopping is only possible if a Raman laser supplies the missing energy; detuning and recoil give the same effects as in the case with different internal states.



\begin{thebibliography}{99}
\bibitem{raman}
 If the coupling involves a multi-photon Raman transition, then the wave-vector $q$ is the appropriate sum/difference of the wave-vectors of each of the lasers, corresponding to the recoil momentum associated with the transition.  Colinear Raman beams yield $q\sim0$.
\bibitem{rf} To achieve large analog magnetic fields, at least one transition must be driven by Raman lasers.
\end{thebibliography}
\end{document}